\documentclass[a4paper,11pt]{article}
\usepackage{pos}
\usepackage{caption}
\usepackage{subcaption}
\usepackage{siunitx}
\DeclareSIUnit{\neq}{n_{eq}}
\DeclareSIUnit{\el}{e^-}
\renewcommand{\logo}{\relax} 

\title{Charge collection and efficiency measurements of the TJ-Monopix2 DMAPS in 180\,nm CMOS technology}
\ShortTitle{TJ-Monopix2: DMAPS in~180\,nm CMOS technology}

\author*[a]{Christian Bespin}
\author[a]{Ivan Caicedo}
\author[a]{Jochen Dingfelder}
\author[a, e]{Tomasz Hemperek}
\author[a, b]{Toko Hirono}
\author[a]{Fabian H\"ugging}
\author[a]{Hans Kr\"uger}
\author[a, c]{Konstantinos Moustakas}
\author[d]{Heinz Pernegger}
\author[d]{Petra Riedler}
\author[a]{Lars Schall}
\author[d]{Walter Snoeys}
\author[a]{Norbert Wermes}

\affiliation[a]{Physikalisches Institut, Universit\"at Bonn,\\
  Nußallee 12, Bonn, Germany}

\affiliation[b]{Deutsches Elektronen-Synchrotron (DESY)\\
Notkesta\ss e. 85, Hamburg, Germany}

\affiliation[c]{Paul Scherrer Institut,\\
Forschungsstrasse 111, Villingen, Switzerland}

\affiliation[d]{CERN\\
Espl. des Particules 1, Meyrin, Switzerland}

\affiliation[e]{DECTRIS AG\\
T\"afernweg 1, Baden-D\"attwil, Switzerland}

\emailAdd{bespin@physik.uni-bonn.de}

\abstract{
  Monolithic CMOS pixel detectors have emerged as competitive contenders in the field of high-energy particle physics detectors. The use of commercial processes offers high-volume production of such detectors.
  A series of prototypes has been designed in a~\SI{180}{\nano\meter} Tower CMOS process with depletion of the sensor material and a column-drain readout architecture.
  The latest iteration, TJ-Monopix2, features a large \SI[product-units=power,parse-numbers=false]{2 \times 2}{\centi\meter\squared} matrix consisting of \num[parse-numbers=false]{512\times 512} pixels with~\SI{33.04}{\micro\meter} pitch.
  A small collection electrode design aims at low power consumption and low noise while the radiation tolerance for high-energy particle detector applications needs extra attention.
  With a goal to reach radiation tolerance to levels of NIEL damage of~\SI{1e15}{1\,\mega\eV\,\neq\per\centi\meter\squared}, a modification of the standard process has been implemented by adding a low-dosed n-type silicon implant across the pixel in order to allow for homogeneous depletion of the sensor volume.
  Recent lab measurements and beam tests were conducted for unirradiated modules to study electrical characteristics and hit detection efficiency.
}

\FullConference{%
  10th International Workshop on Semiconductor Pixel Detectors for Particles and Imaging (Pixel2022)\\
  12-16 December 2022\\
  Santa Fe, New Mexico, USA}

\begin{document}
\maketitle

\section{Introduction}

In recent years, advances in CMOS technologies have fueled the development of a new generation of monolithic active pixel sensors (MAPS) with fast readout and high radiation tolerance by depleting the charge sensitive volume~\cite{peric_hvcmos}.
These depleted MAPS (DMAPS) devices are therefore an interesting candidate for high-energy particle physics experiments with high radiation environments and high particle rate.
Depletion is achieved by either using high-voltage add-ons in the CMOS technology and/or high resistivity substrates.
The increasing availability of these features in commercial CMOS processes could combine the features of the detector concept with possibly faster and cheaper production than common hybrid pixel detectors for the mentioned purposes.
The idea behind and measurements results from one of multiple DMAPS prototypes, TJ-Monopix2~\cite{moustakas_cmos_design, moustakas_thesis}, will be presented in the following.

\section{Design of TJ-Monopix2}

\begin{figure}[ht]
  \centering
  \begin{subfigure}[t]{0.48\textwidth}
    \centering
    \includegraphics[width=0.97\textwidth,trim={0 0.2cm 0 -0.2cm},clip]{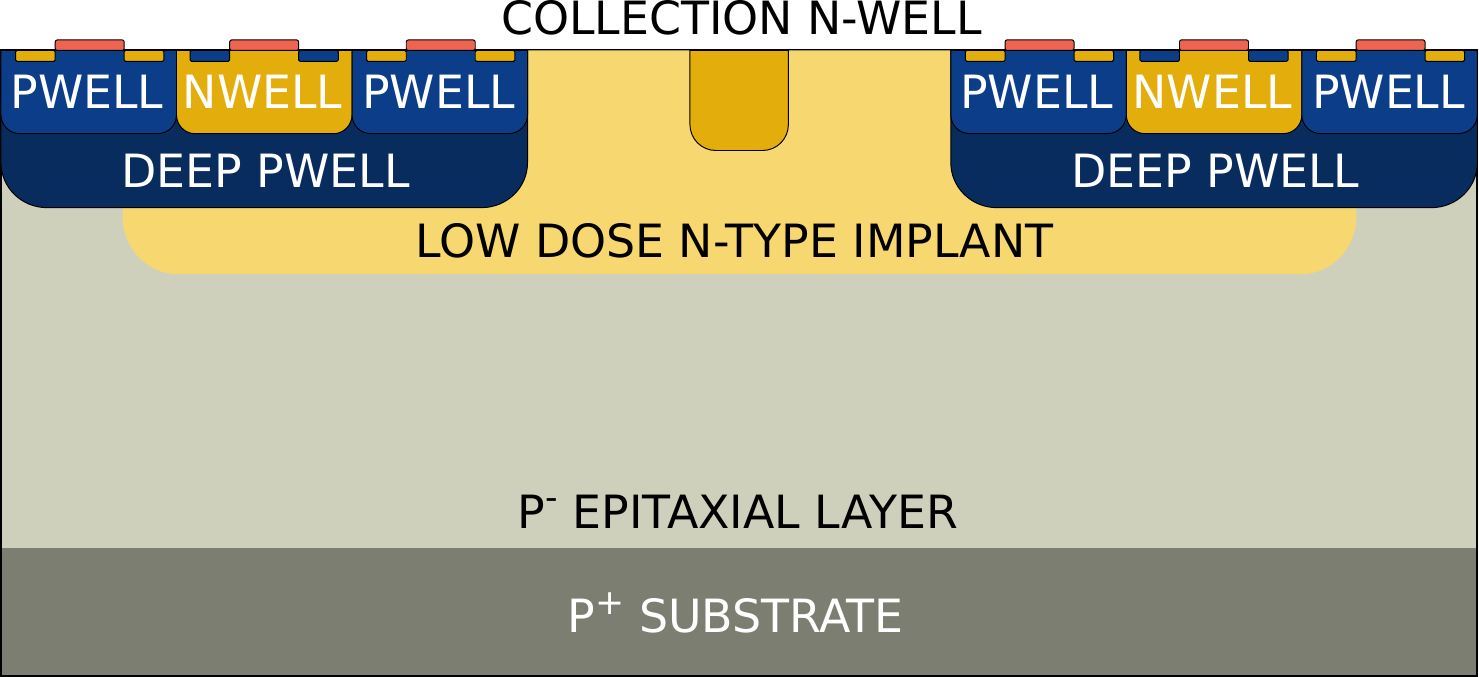}
    \caption{Modification with gap in low dose n-type implant below pixel electronics.}
    \label{subfig:ngap}
  \end{subfigure}
  \begin{subfigure}[t]{0.48\textwidth}
    \centering
    \includegraphics[width=0.97\textwidth,trim={0 0.2cm 0 -0.2cm},clip]{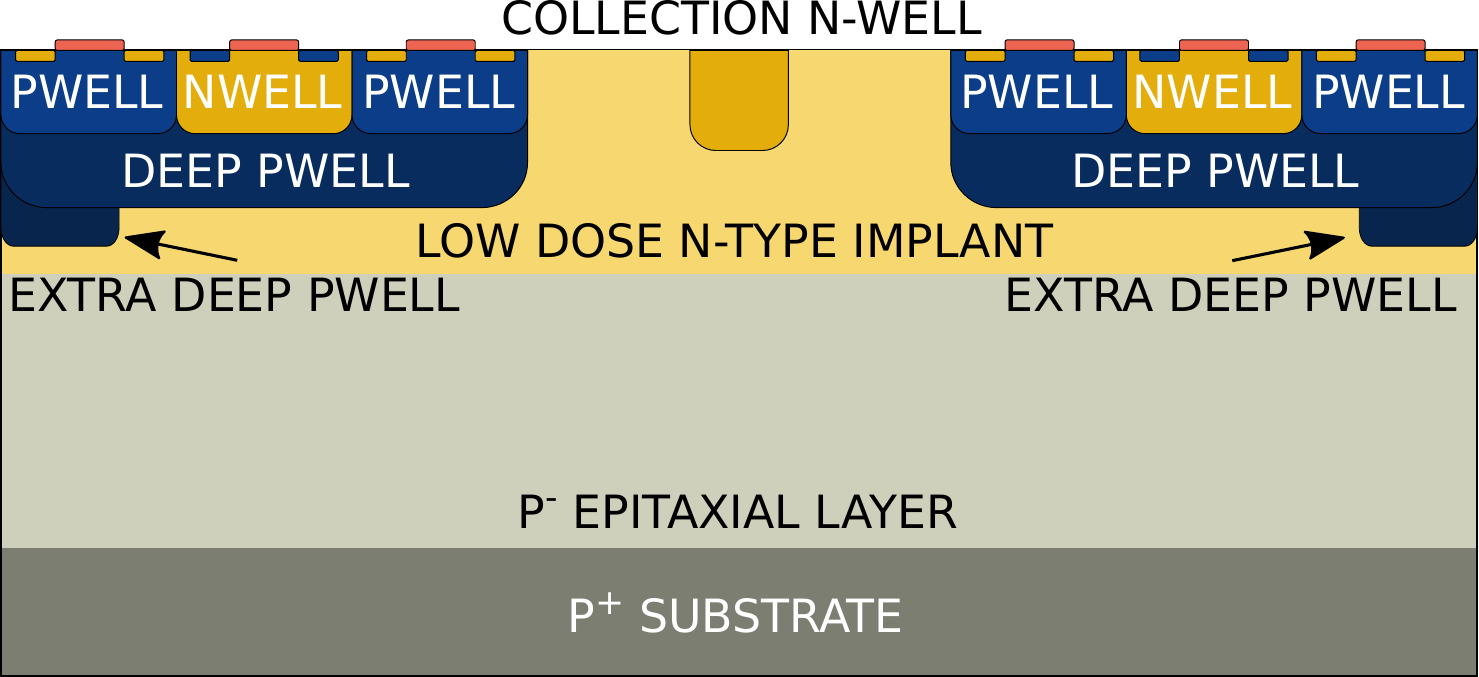}
    \caption{Modification with continuous n-type implant and deep p-type implant below pixel electronics.}
    \label{subfig:xdpw}
  \end{subfigure}
  \caption{Cross-section variants of modified sensor process for TJ-Monopix2.}
  \label{fig:crosssection}
\end{figure}

TJ-Monopix2 is the latest DMAPS prototype from the TJ-Monopix development line which is based on the ALPIDE pixel detector developed for the ALICE ITS upgrade~\cite{mager_alpide}.
It is fabricated in the same~\SI{180}{\nano\meter} commercial CMOS process provided by Tower Semiconductor\footnote{\url{https://towersemi.com}}.
A modification of the process used for ALPIDE has been implemented to increase the radiation tolerance to levels~$\geq\SI{1e15}{\neq\per\centi\meter\squared}$ by adding a low dose n-type implant for homogeneous growth of the depletion zone with applied bias voltage.
In measurements on first prototypes with this modification, a drop in hit detection efficiency was observed after irradiation~\cite{caicedo_monopix1, bespin_dmaps}.
This could be improved significantly by adding a gap in the n-type blanket or a deep p-type implant in the pixel corners to shape the electrical field towards the collection electrode~\cite{dyndal_minimalta}.
The cross-sections of these two sensor designs is shown in fig.~\ref{fig:crosssection}.
Additionally, chips have been produced on Czochralski silicon to increase the available depletable volume compared to the thickness of the epitaxial layer ($\mathcal{O}\!\left(\SI{10}{\micro\meter}\right)$).
Measurements on Czochralski silicon chips in TJ-Monopix1 showed a further increase in hit detection efficiency after irradiation~\cite{Bespin2022}.

TJ-Monopix2 follows a small collection electrode approach with a pixel capacitance of about~\SI{3}{\femto\farad}.
The pixels of size~\SI[parse-numbers=false]{33\times 33}{\micro\meter\squared} are read out using an established synchronous column-drain technique from the FE-I3 readout chip~\cite{peric_fei3}.
Further changes from the predecessor TJ-Monopix1 include an improved front-end design, a new pixel masking scheme and a 3-bit DAC for local threshold tuning.
With these changes the threshold is expected to be reduced by a factor of~\num{3} while improving the threshold dispersion and noise behavior.

The digital periphery contains logic for register configuration, data handling and LVDS output drivers.
Slow control is done via a command protocol and decoder that was taken from the RD53B readout chip~\cite{rd53b_manual}.
Both pixel and register data is~8b10b encoded in a frame-based data stream which allows operating the chip with four differential data lines.

\section{Injection-based threshold and noise measurements}

\begin{figure}[ht]
  \centering
  \begin{subfigure}[t]{0.49\textwidth}
    \centering
    \includegraphics[width=0.97\textwidth,trim={0.4cm 0.3cm 0.6cm 1.35cm},clip]{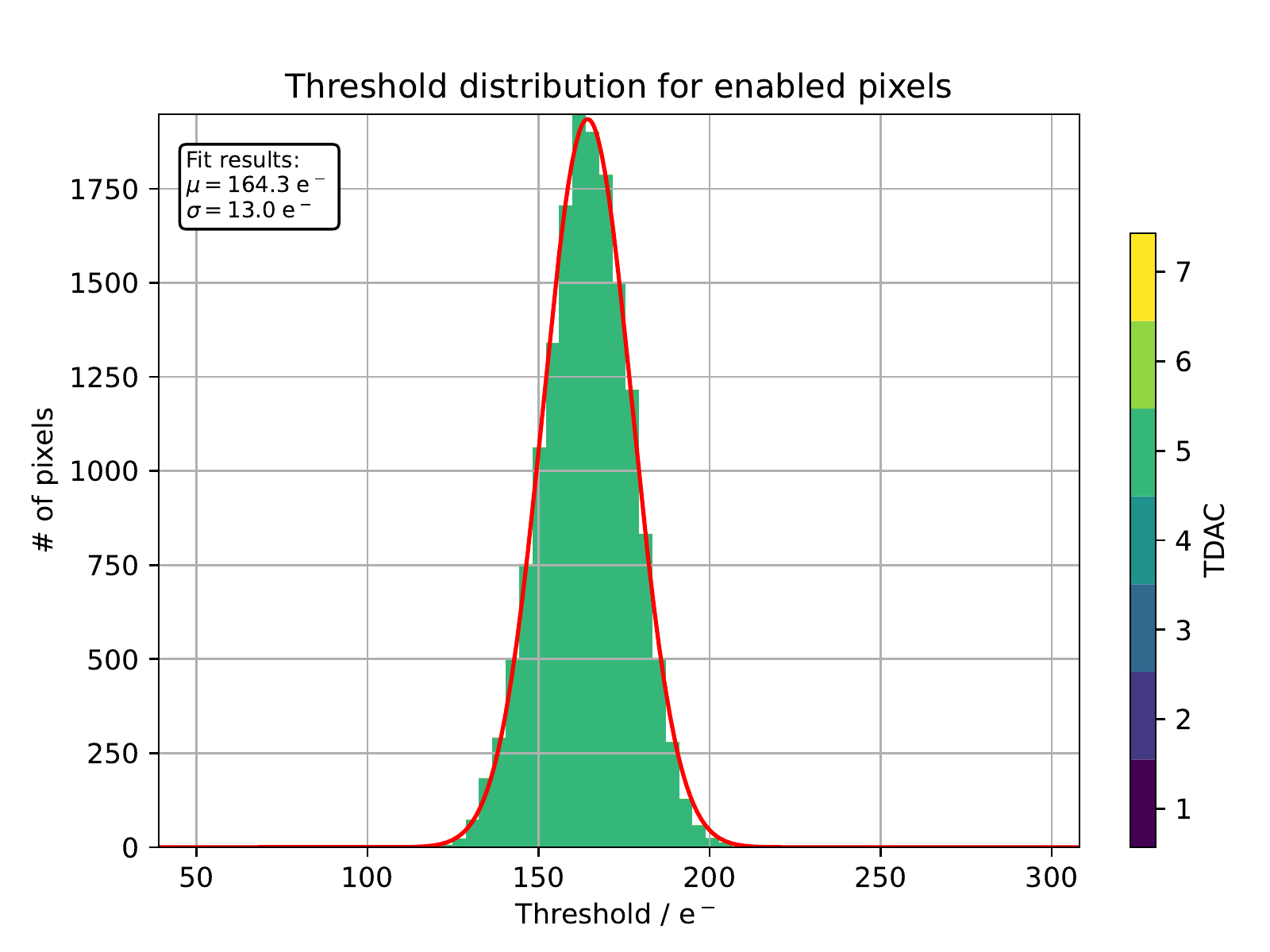}
    \caption{Threshold distribution before in-pixel threshold tuning.}
    \label{fig:thr_disp_untuned}
  \end{subfigure}
  \begin{subfigure}[t]{0.49\textwidth}
    \centering
    \includegraphics[width=0.97\textwidth,trim={0.4cm 0.3cm 0.6cm 1.35cm},clip]{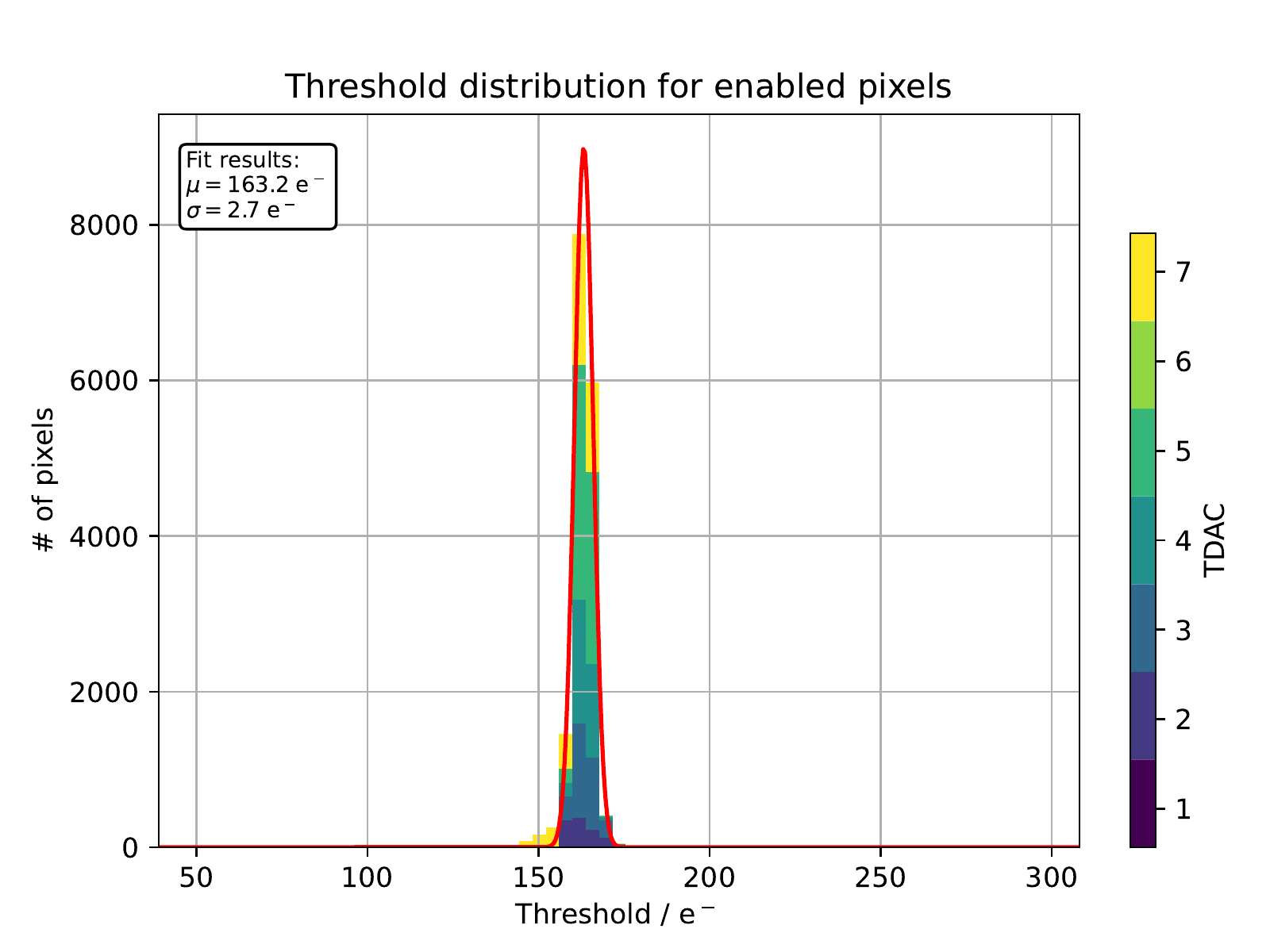}
    \caption{Threshold distribution after in-pixel threshold tuning.}
    \label{fig:thr_disp_tuned}
  \end{subfigure}
  \caption{Threshold distribution of TJ-Monopix2 before and after adjusting the in-pixel threshold DAC to lower the dispersion. Color-coded is the value of the DAC setting for every pixel.}
  \label{fig:thr_tuning}
\end{figure}

Initial tests have been performed in a laboratory setup to measure the threshold and noise performance of TJ-Monopix2.
All of these values are extracted from injecting different amounts of charge into the pixel a given number of times and recording the amount of registered hits $n_\mathrm{hits}$. 
The response function is a smeared step function of the form
\begin{align}
  n_\mathrm{hits}(q) = \dfrac{1}{2} \cdot n_\mathrm{injections} \cdot \left(\mathrm{erf}\left(\dfrac{q - q_\mathrm{thr}}{\sigma\,\sqrt{2}}\right) + 1\right)
  \label{eqn:s_curve}
\end{align}
with $q$ the injected charge amount, $n_\mathrm{injections}$ the number of consecutive injections of $q$ and $q_\mathrm{thr}$ the charge at the threshold.
$\sigma$ denotes the gaussian smearing of the step function which is defined as the electronic noise of the pixel.
The threshold is defined as the charge at which~\SI{50}{\percent} of the injected hits are recorded from the pixel.
Histogramming the individual thresholds from each pixel leads to the distribution shown in fig.~\ref{fig:thr_disp_untuned}.
The distribution has a mean value of~\SI{164}{\el} and a width of~\SI{13}{\el} which is defined as the threshold dispersion.
By adjusting the threshold DAC in each pixel in order to even out the deviations from the target threshold, the threshold dispersion can be reduced significantly.
The resulting distribution after this so-called tuning process is shown in fig.~\ref{fig:thr_disp_tuned}.
While the mean threshold stays basically the same, the dispersion could be reduced by a factor of almost~\num{5}.
Both the mean threshold and threshold dispersion are significantly lower than in TJ-Monopix1, where losses in hit detection efficiency could be observed due to too large thresholds~\cite{Bespin2022}.

\begin{figure}[ht]
  \centering
  \centering
  \begin{subfigure}[t]{0.49\textwidth}
    \centering
    \includegraphics[width=0.97\textwidth,trim={0.4cm 0.3cm 0.6cm 1.35cm},clip]{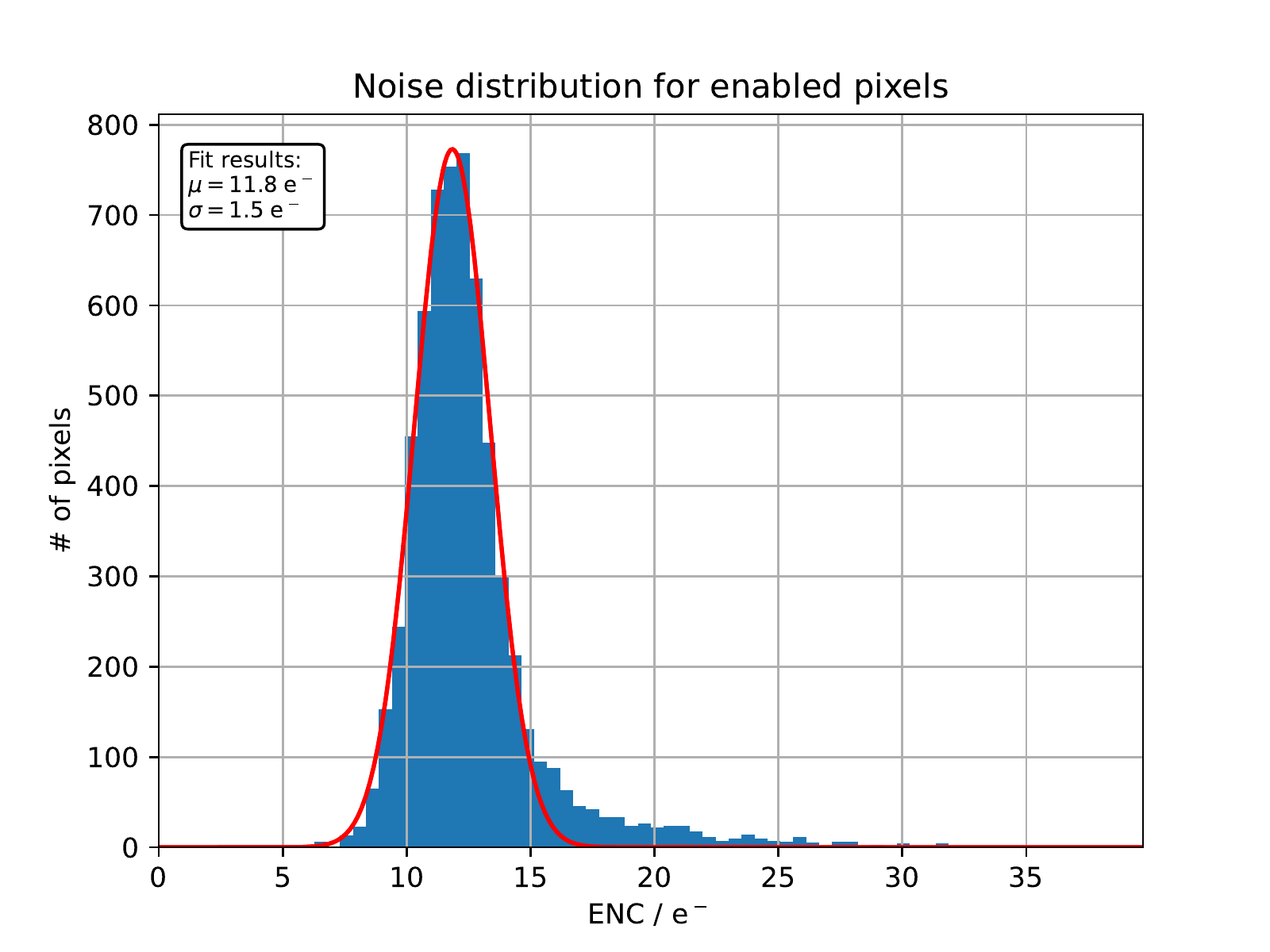}
    \caption{Noise distribution of TJ-Monopix1 with noticeable tail towards larger values.}
    \label{fig:enc_tj1}
  \end{subfigure}
  \begin{subfigure}[t]{0.49\textwidth}
    \centering
    \includegraphics[width=0.97\textwidth,trim={0.4cm 0.3cm 0.6cm 1.35cm},clip]{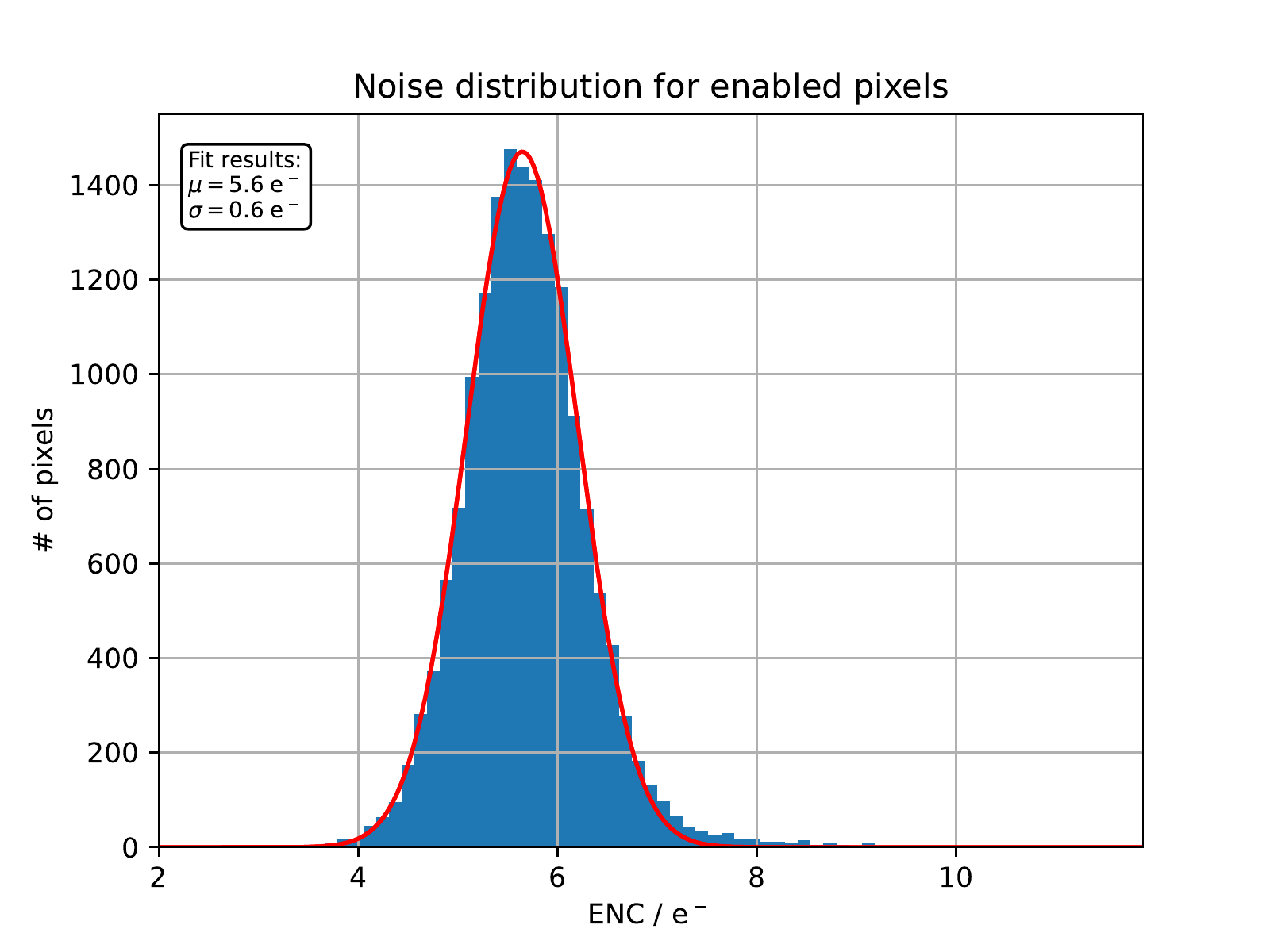}
    \caption{Noise distribution of TJ-Monopix2. There is no observable tail and lower noise overall.}
    \label{fig:enc_tj2}
  \end{subfigure}
  \caption{Noise distribution of TJ-Monopix1 (left) and TJ-Monopix2 (right) for comparison.}
  \label{fig:enc}
\end{figure}

The corresponding histogram of the electronic noise is depicted in fig.~\ref{fig:enc}.
As a comparison, the distribution from the predecessor TJ-Monopix1 is included, where a large tail towards higher values was observed that led to a high operational threshold in order to limit the amount of noisy pixels.
It can be seen, that this tail is largely removed with slight changes to the analog front-end, which in turn lowers the threshold for a regular operation of the chip.

\section{Hit detection efficiency measurements}

Two different pixel variations were investigated regarding their hit detection efficiency, that will be presented in the following -- a DC-coupled, more standard design which makes up most part of the matrix and an AC-coupled investigative design realized in only a few columns of the matrix.
While the former was measured in more detail, some first results of the latter are included as well.

\subsection{Standard DC-coupled pixel flavor}

First measurements to determine the hit detection efficiency have been performed in a ~\SI{5}{\giga\electronvolt} electron beam at the DESY II testbeam facility at DESY, Hamburg~\cite{desy_testbeam}.
Three unirradiated modules were tested with different sensor geometries: two chips with~\SI{30}{\micro\meter} thick epitaxial silicon and the two geometries depicted in fig.~\ref{fig:crosssection} as well as one chip built on~\SI{300}{\micro\meter} Czochralski silicon with a gap in the low dose n-type implant (see fig.~\ref{subfig:ngap}).
It should be noted that the different substrate materials offer different sensor thicknesses and therefore charge-sensitive volume depending on the depletion.
The measurements are not targeting a comparison between different types of silicon.

\begin{figure}[ht]
  \centering
  \begin{subfigure}[t]{0.49\textwidth}
    \centering
    \includegraphics[width=0.97\textwidth,trim={0.4cm 0.3cm 0.6cm 1cm},clip]{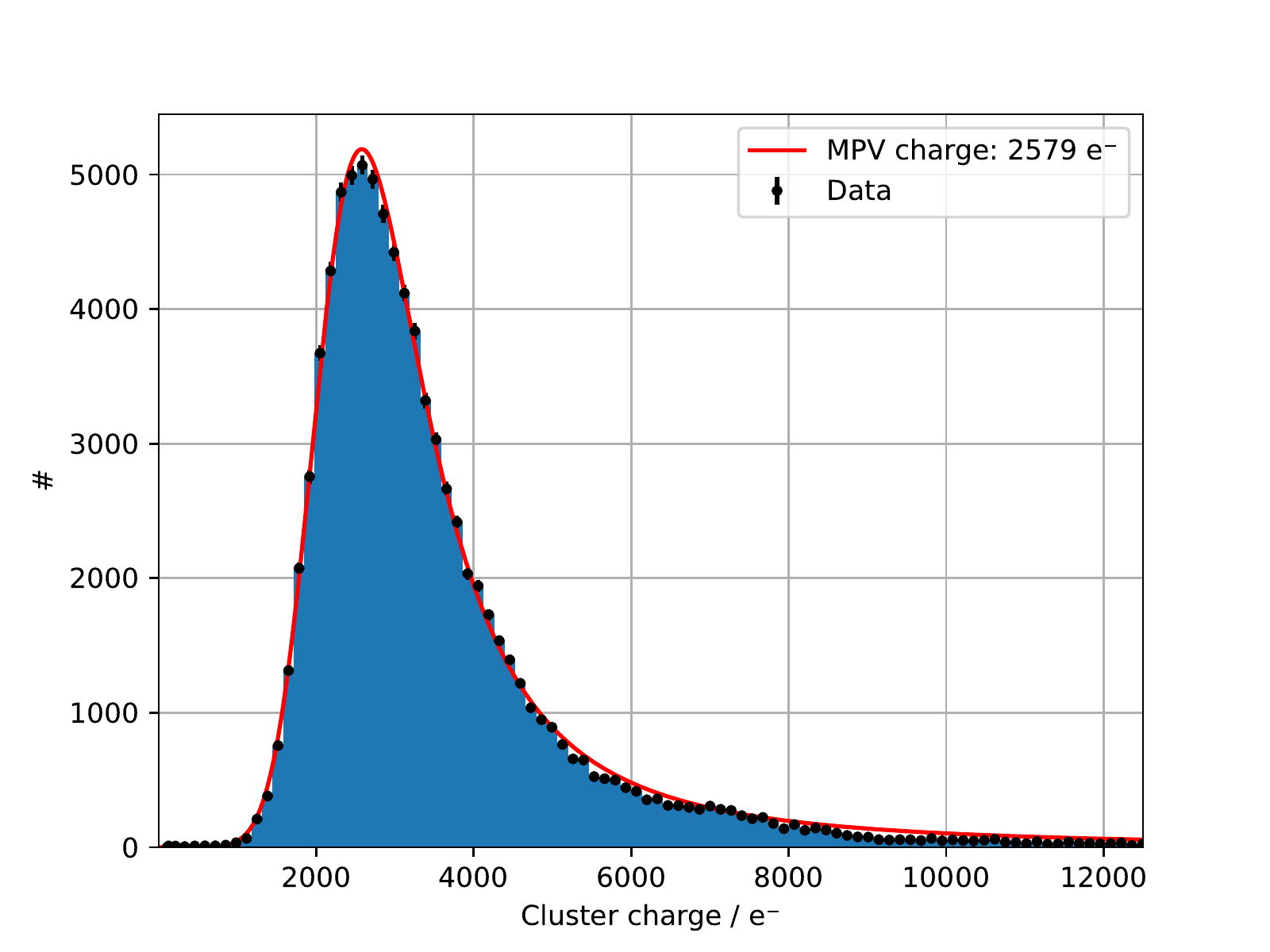}
    \caption{Cluster charge distribution for an epitaxial silicon chip.}
    \label{fig:cl_charge_epi}
  \end{subfigure}
  \begin{subfigure}[t]{0.49\textwidth}
    \centering
    \includegraphics[width=0.97\textwidth,trim={0.4cm 0.3cm 0.6cm 1cm},clip]{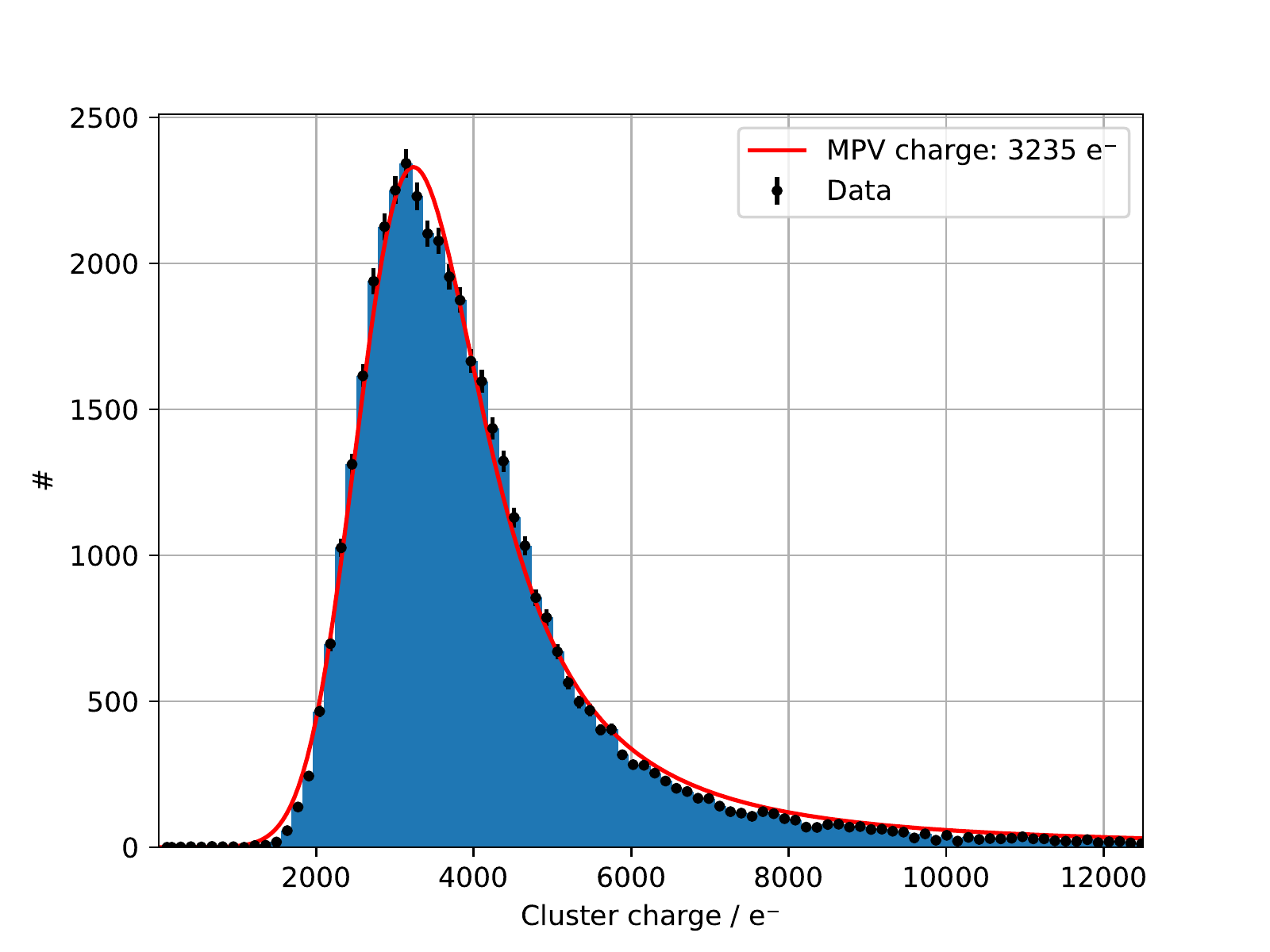}
    \caption{Cluster charge for a Czochralski silicon chip.}
    \label{fig:cl_charge_cz}
  \end{subfigure}
  \begin{subfigure}[t]{0.49\textwidth}
    \centering
    \includegraphics[width=0.97\textwidth,trim={0.4cm 0.3cm 0.6cm 0.5cm},clip]{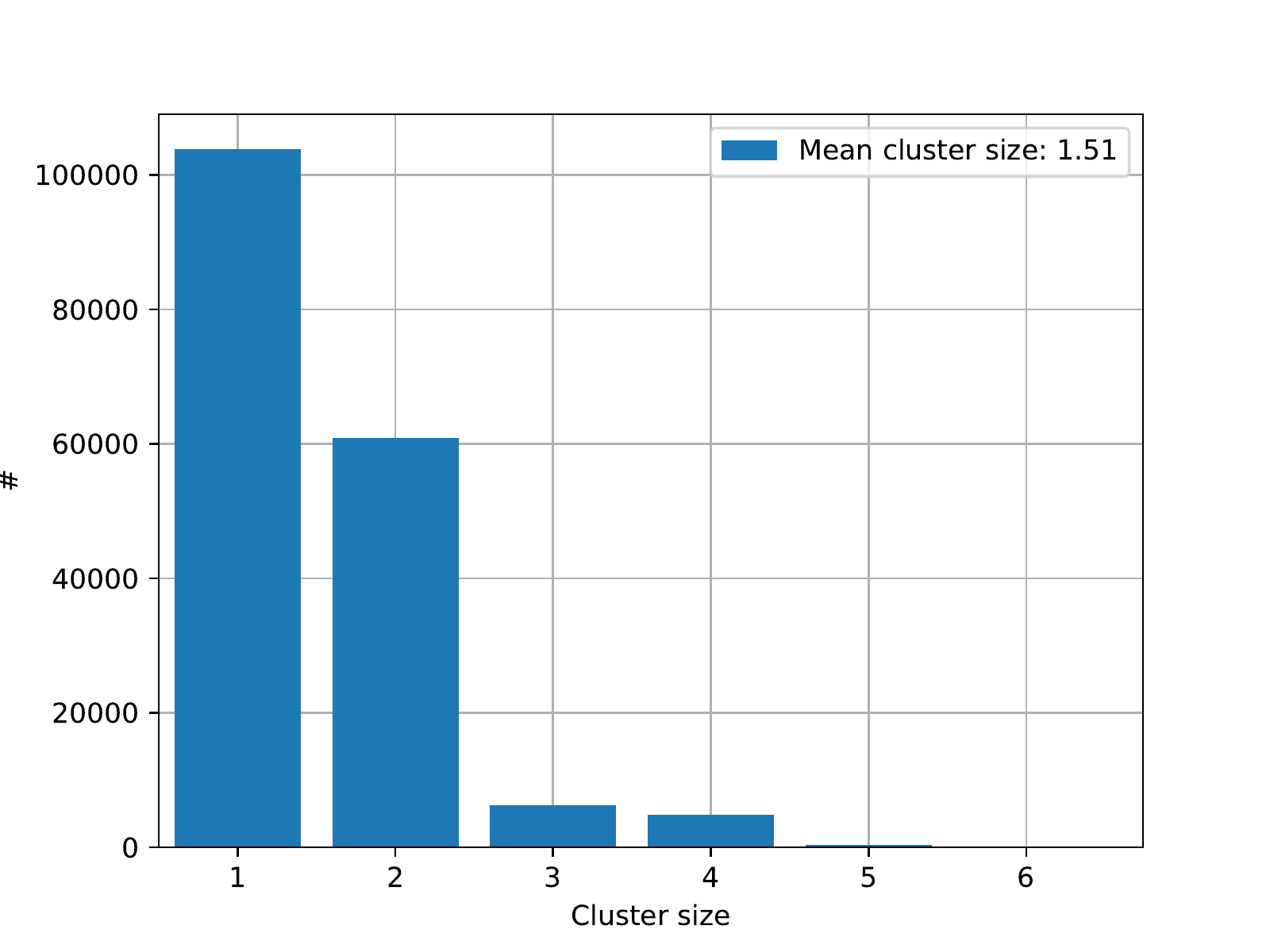}
    \caption{Cluster size distribution for an epitaxial silicon chip.}
    \label{fig:cl_size_epi}
  \end{subfigure}
  \begin{subfigure}[t]{0.49\textwidth}
    \centering
    \includegraphics[width=0.97\textwidth,trim={0.4cm 0.3cm 0.6cm 0.5cm},clip]{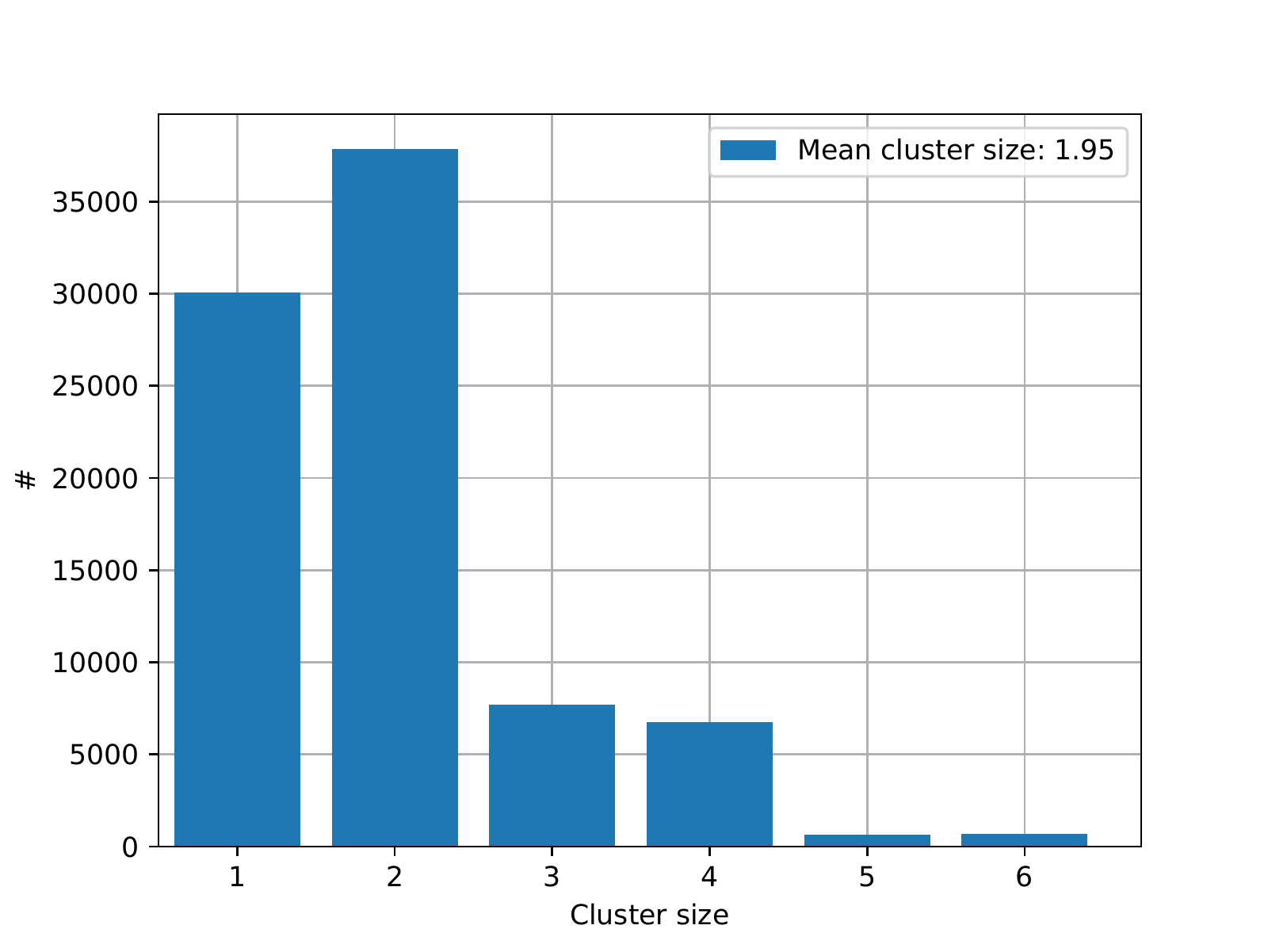}
    \caption{Cluster size distribution for a Czochralski silicon chip.}
    \label{fig:cl_size_cz}
  \end{subfigure}
  \caption{Cluster charge and size distributions for a chip with~\SI{30}{\micro\meter} epitaxial silicon (left) and~\SI{300}{\micro\meter} Czochralski silicon (right) at~\SI{-6}{\volt} bias voltage. The latter can be depleted further than~\SI{30}{\micro\meter} resulting in a larger cluster charge and size. Both chips were operated at a threshold of~\SI{200}{\el}.}
  \label{fig:clusters}
\end{figure}

Figures~\ref{fig:cl_charge_epi} and~\ref{fig:cl_charge_cz} show the recorded cluster charge for a chip with epitaxial layer and with Czochralski substrate.
It can be observed that the collected charge is about~\SI{25}{\percent} larger in the Cz sample, because the depletion depth is only limited by the thickness of the sensor (\SI{300}{\micro\meter}) which is by far not fully depleted, but more depleted than the~\SI{30}{\micro\meter} thick epitaxial layer in the other chip.

The average cluster size is significantly larger in the Cz sample as well which results in a high spatial resolution due to charge-weighted clustering. 
The cluster size distributions for the same samples as above are depicted in figs.~\ref{fig:cl_size_epi} and~\ref{fig:cl_size_cz}.
While cluster size~\num{1} is predominant in the epitaxial sample, the Cz sample has mainly clusters of size~\num{2}. The corresponding average cluster size is~\num{1.55} and~\num{1.95}, respectively.
Taking the pointing resolution of the beam telescope into account, an intrinsic spatial resolution of~\SI{8.6}{\micro\meter} could be achieved in a Czochralski silicon sample.

\begin{figure}[ht]
  \centering
  \begin{subfigure}[t]{0.48\textwidth}
    \centering
    \vspace{1em}
    \includegraphics[width=0.95\textwidth,trim={0 0.3cm 0 1.2cm},clip]{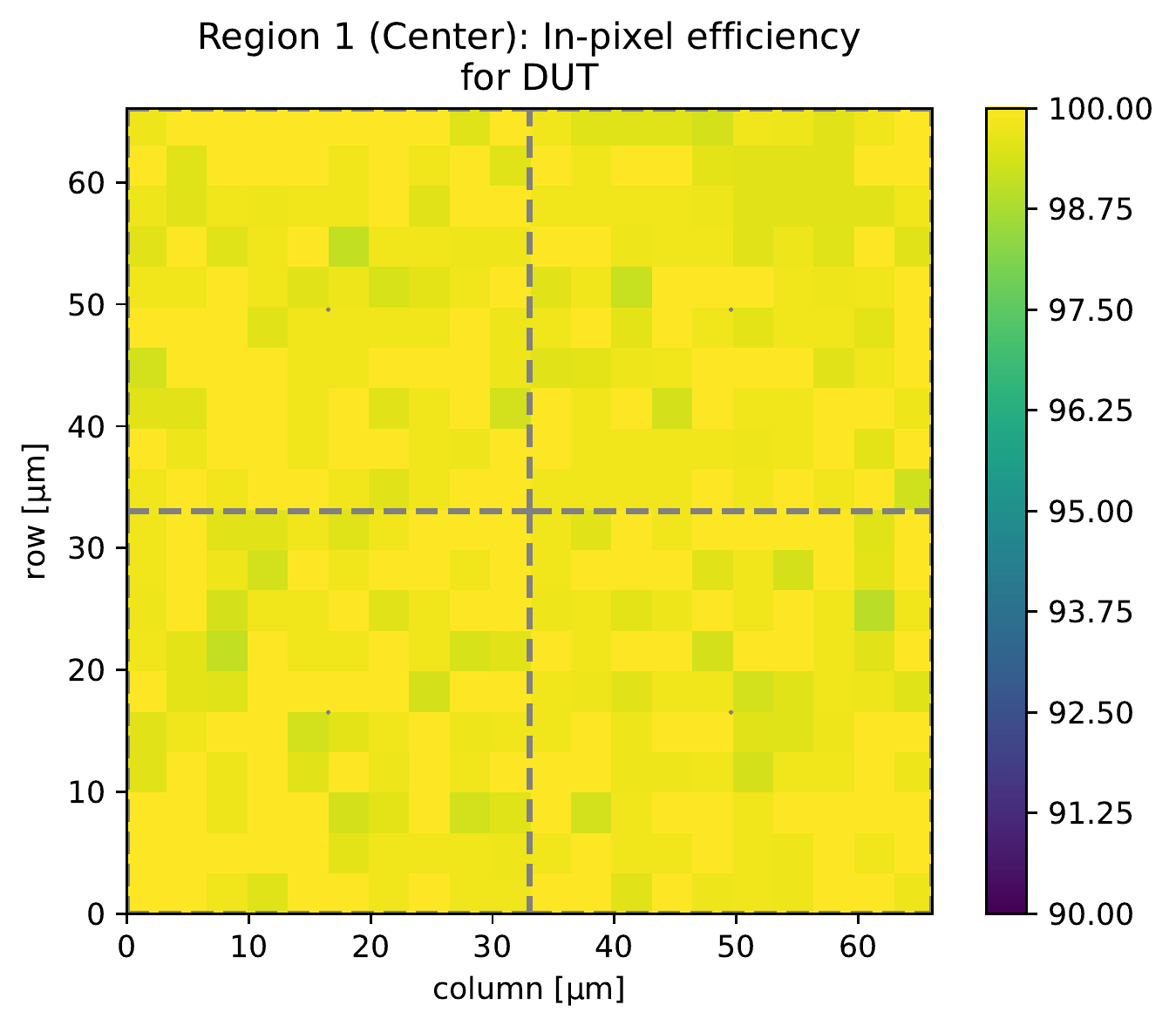}
    \caption{\SI{99.80\pm0.1}{\percent} efficiency for a chip built on epitaxial silicon with gap in n-layer modification from fig.~\ref{subfig:ngap}.}
    \label{fig:eff_epi_ngap}
  \end{subfigure}
  \begin{subfigure}[t]{0.48\textwidth}
    \centering
    \vspace{1em}
    \includegraphics[width=0.95\textwidth,trim={0 0.3cm 0 1.2cm},clip]{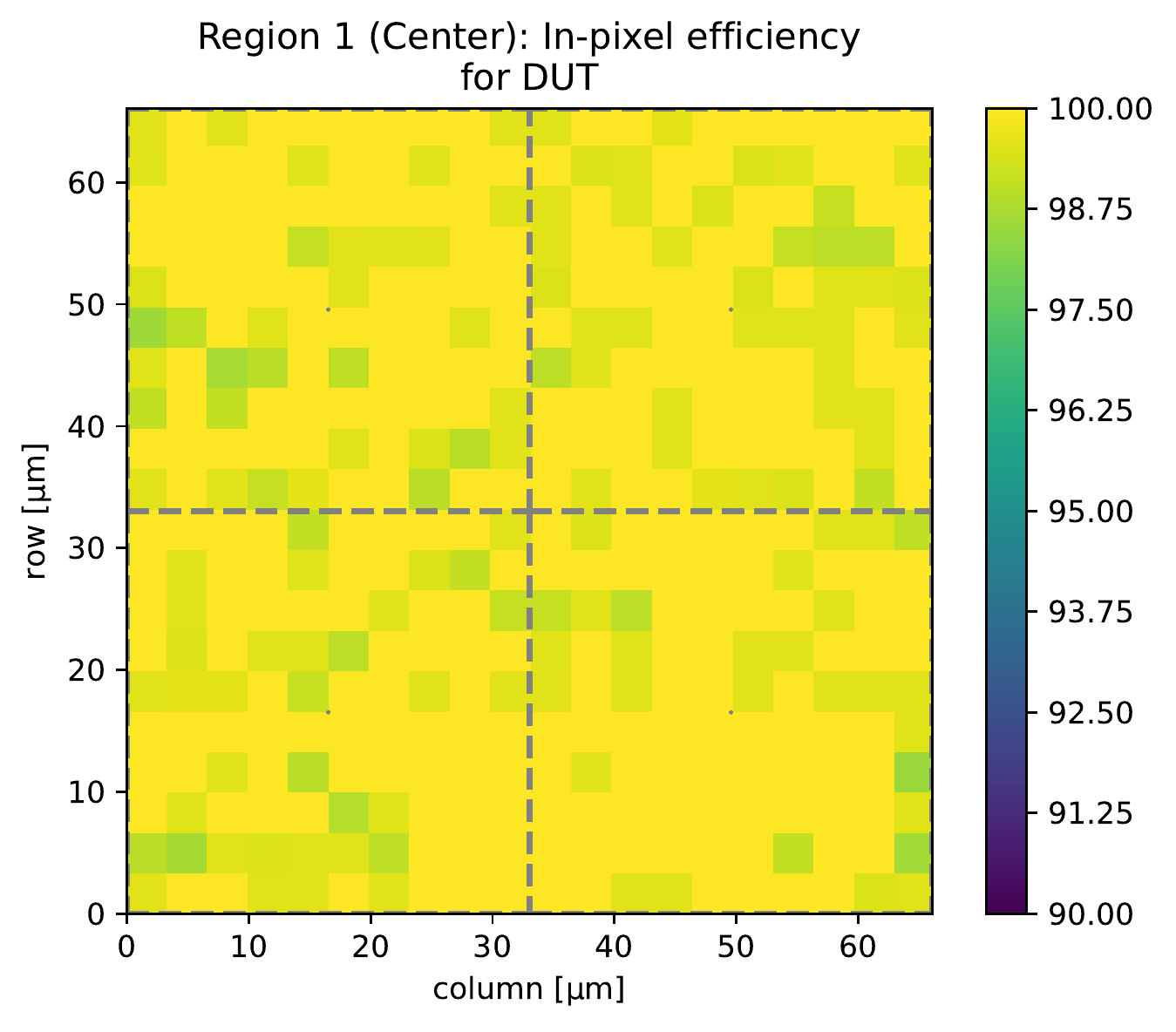}
    \caption{\SI{99.79\pm0.1}{\percent} efficiency for a chip built on Cz silicon with gap in n-layer modification from fig.~\ref{subfig:ngap}.}
    \label{fig:eff_cz_ngap}
  \end{subfigure}
  \begin{subfigure}[t]{0.48\textwidth}
    \centering
    \vspace{1em}
    \includegraphics[width=0.95\textwidth,trim={0 0.3cm 0 1.2cm},clip]{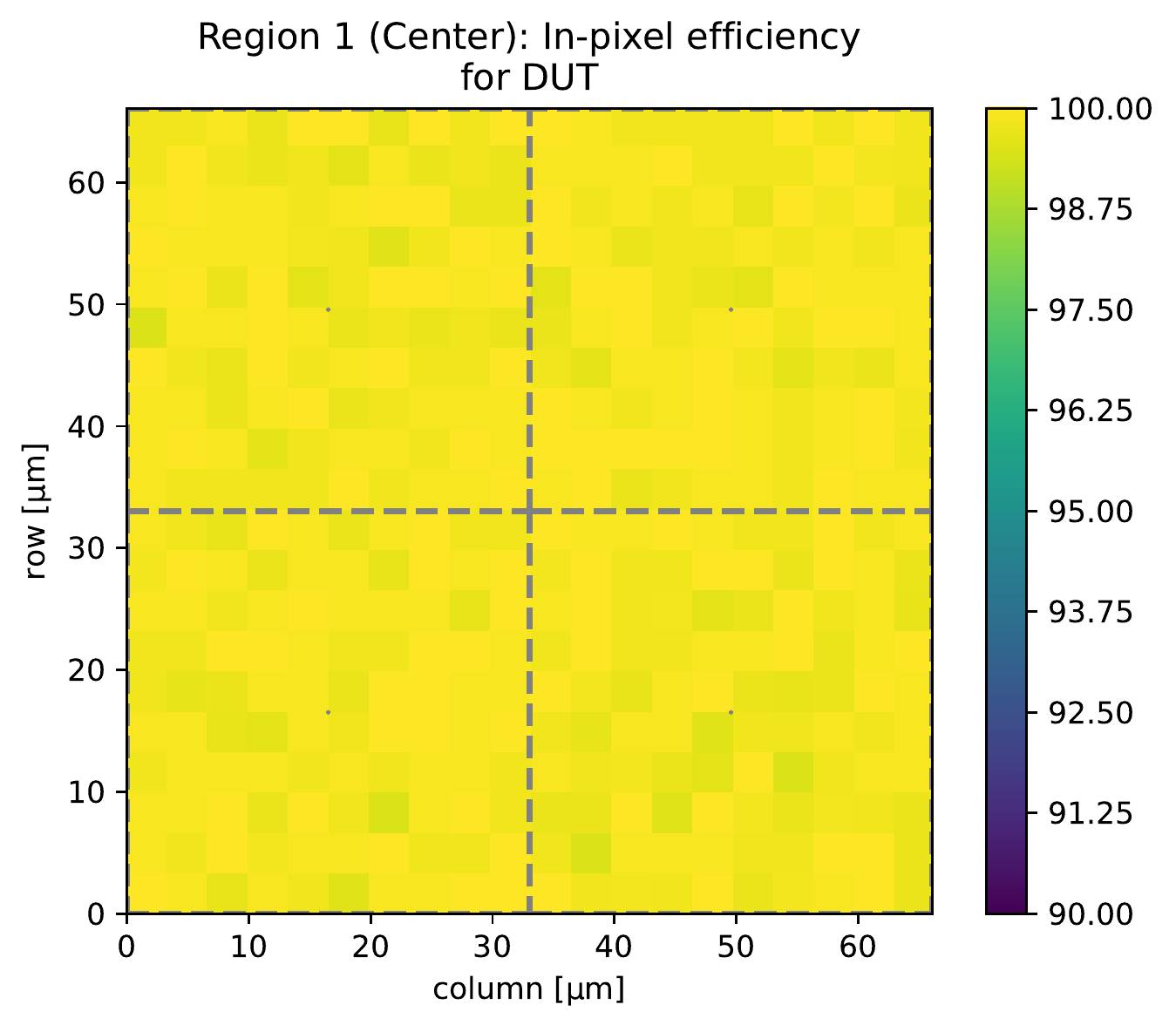}
    \caption{\SI{99.85\pm0.1}{\percent} efficiency for a chip built on epitaxial silicon with additional p-well modification from fig.~\ref{subfig:xdpw}.}
    \label{fig:eff_epi_xdpw}
  \end{subfigure}
  \caption{Hit detection efficiencies for different substrate materials with different thicknesses and sensor geometries. Results were mapped onto a~2~x~2 pixel array for higher statistics and in-pixel resolution. The chips were operated with~\SI{-6}{\volt} bias voltage and at a~\SI{200}{\el} threshold.}
  \label{fig:eff_normal}
\end{figure}

The hit detection efficiency was measured with a beam telescope with six \textsc{Mimosa26} planes and a \textsc{FE-I4} time reference plane which are all connected to a trigger logic unit to synchronize individual detector hits time-wise.
The efficiency for all three modules is shown in fig.~\ref{fig:eff_normal} where the result for every pixel was mapped onto a two by two pixel cell to increase the statistics to see possible effects or efficiency losses within a single pixel.
All samples were running at a threshold of about~\SI{200}{\el} and achieve a hit detection efficiency around~\SI{99.80}{\percent} with slight deviations within the error (estimated $< \SI{0.1}{\percent}$).
There are no losses observable in the pixel corners or between pixels.

\subsection{AC-coupled pixel flavor}

\begin{figure}[ht]
  \centering
  \begin{subfigure}[t]{0.48\textwidth}
    \centering
    \includegraphics[width=0.95\textwidth,trim={0 0 0 1.1cm},clip]{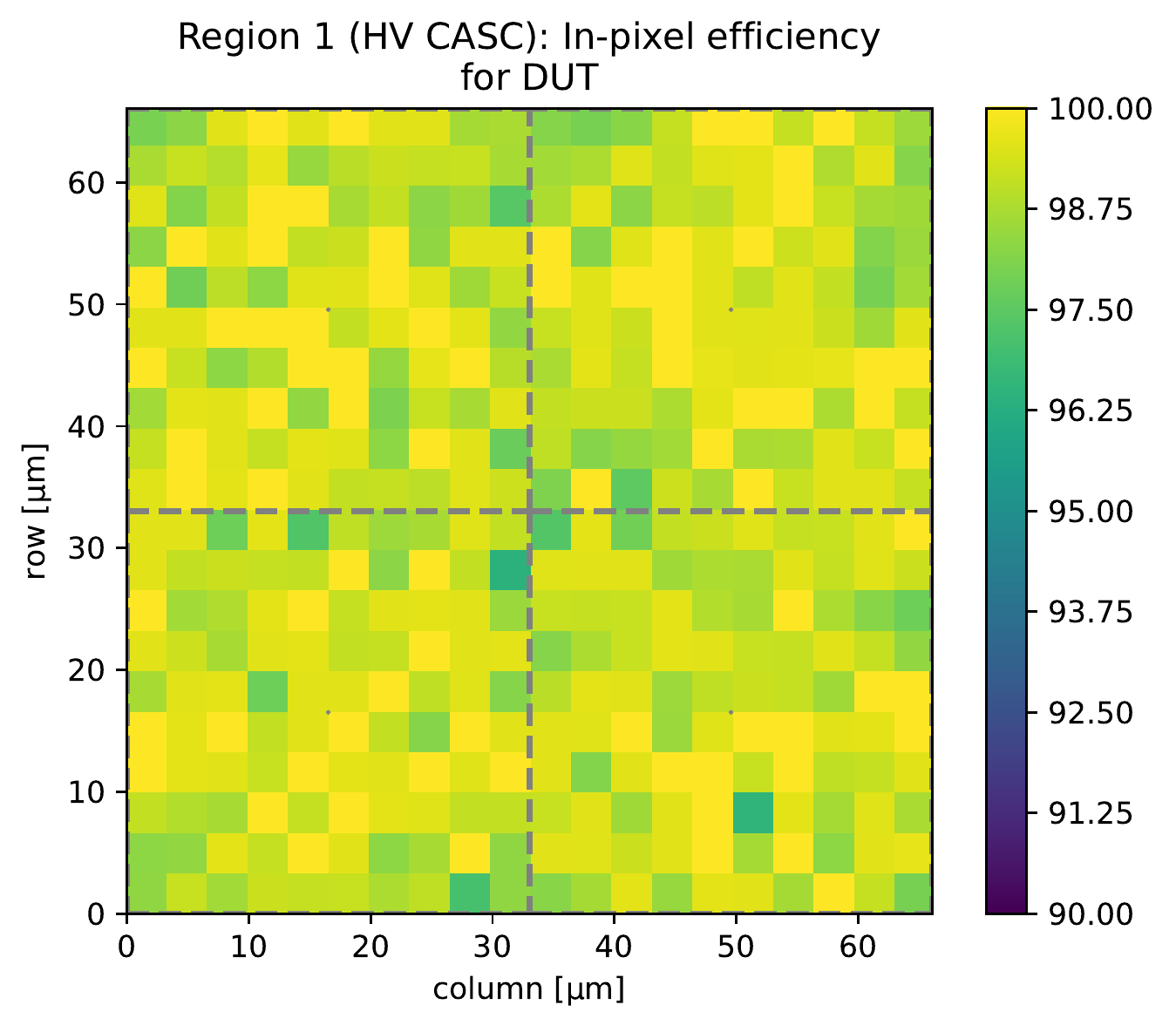}
    \caption{\SI{99.21\pm0.1}{\percent} efficiency of an AC-coupled pixel flavor at~\SI{5}{\volt} bias voltage and~\SI{250}{\el} threshold.}
    \label{fig:eff_hv_5}
  \end{subfigure}
  \begin{subfigure}[t]{0.48\textwidth}
    \centering
    \includegraphics[width=0.95\textwidth,trim={0 0 0 1.1cm},clip]{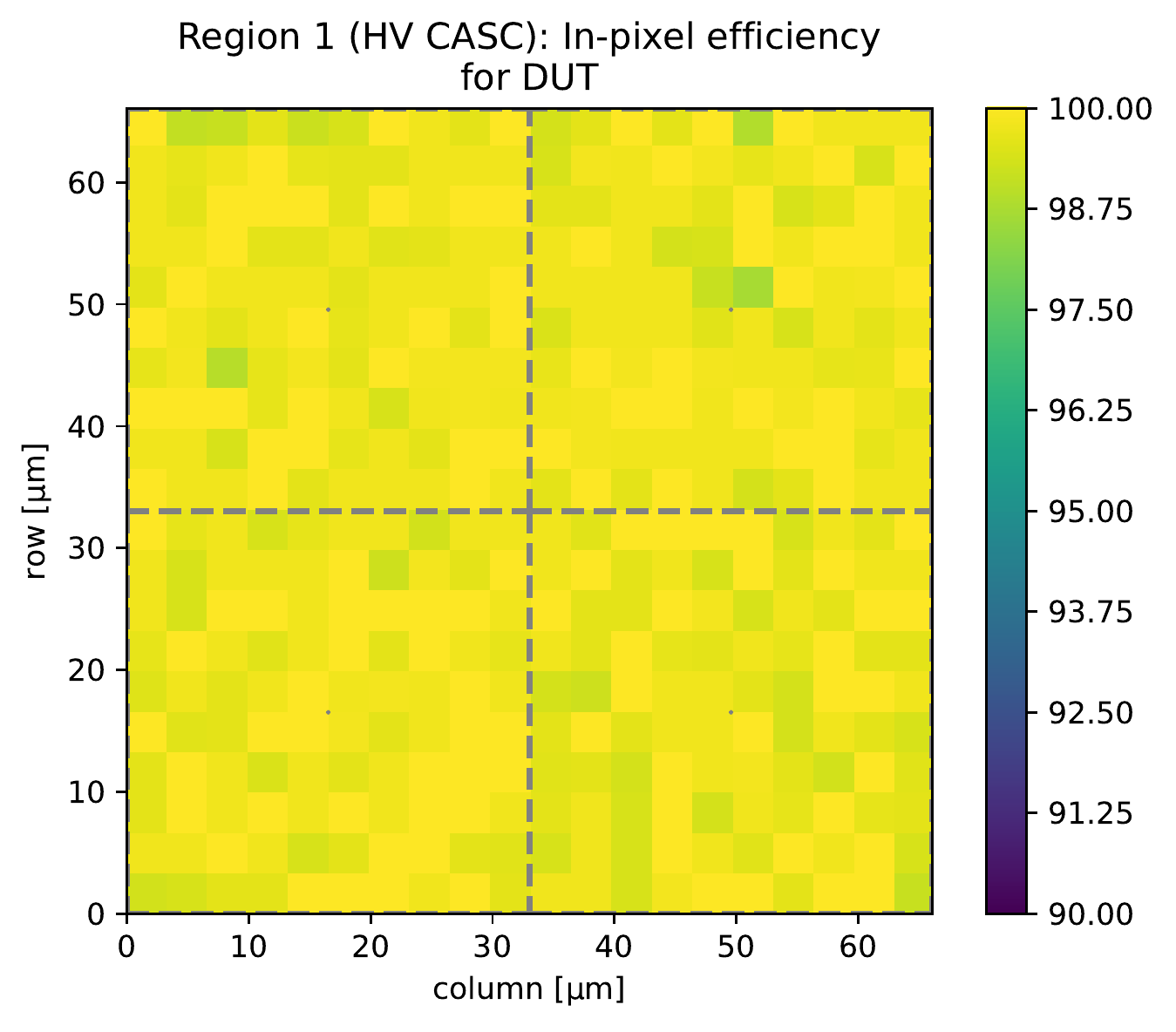}
    \caption{\SI{99.85\pm0.1}{\percent} efficiency of an AC-coupled pixel flavor at~\SI{25}{\volt} bias voltage and~\SI{200}{\el} threshold.}
    \label{fig:eff_hv_25}
  \end{subfigure}
  \caption{Hit detection efficiency of an AC-coupled pixel flavor at~(\ref{fig:eff_hv_5})~\SI{5}{\volt} and~(\ref{fig:eff_hv_25})~\SI{25}{\volt} bias voltage.}
  \label{fig:eff_hv}
\end{figure}

Another pixel variation with different analog front-end was tested as well to determine its performance in a particle beam.
In this design the (positive) bias voltage is applied via a diode on the top side of the chip and connected to the charge collection n-well.
To avoid breakdown of the front-end electronics due to the high voltage ($\leq\SI{50}{\volt}$) on that well, the input signal is AC coupled to the amplifier.
This approach can potentially deplete the substrate further due to the higher voltage than what can be applied in the standard pixel design.
The hit detection efficiency was measured for different bias voltages and is shown in fig.~\ref{fig:eff_hv}.
At~\SI{5}{\volt} the efficiency is already above~\SI{99}{\percent} and reaches the same value as for the DC coupled pixel flavor of~\SI{99.85}{\percent} at or before~\SI{25}{\volt} bias voltage.
This is, taking the slightly higher threshold into account, in agreement with the expectation that there should be no noticeable difference in hit detection efficiency before irradiation between the two pixel flavors.
The larger applicable bias voltage could prove superior after irradiation to achieve more depletion and therefore higher charge signal.

\section{Conclusion}

In summary, the performance of TJ-Monopix2 shows a significant improvement in threshold value and dispersion compared to TJ-Monopix1, although the former is higher than its design value (\SI{120}{\el}).
With the measured amount of charge in the sensor the threshold is still small enough to detect a majority of hits even from large clusters before irradiation.
The large signal compared to the small pixel leads to a large cluster size and therefore high spatial resolution, where chips on Czochralski substrate perform slightly better due to the larger sensor volume.
For the tested sensor materials with different thicknesses and sensor geometries, the hit detection efficiency is around~\SI{99.8}{\percent} or better in all cases.
The modified front-end version with bias applied on the charge collection node achieves similar values for the hit detection efficiency while providing a larger headroom in bias voltage to achieve efficient performance after radiation damage.
The results for irradiated modules will be presented in a forthcoming publication.

\acknowledgments
This project has received funding from the Deutsche Forschungsgemeinschaft DFG (grant WE 976/4-1), the German Federal Ministry of Education and Research BMBF (grant 05H15PDCA9), and the European Union’s Horizon 2020 research and innovation program under grant agreements no. 675587 (Maria Sklodowska-Curie ITN STREAM), 654168 (AIDA-2020), and 101004761 (AIDAinnova).
The measurements leading to these results have been performed at the Test Beam Facility at DESY Hamburg (Germany), a member of the Helmholtz Association (HGF).

\bibliographystyle{jhep}
\bibliography{bibliography}

\end{document}